\documentclass[fleqn,usenatbib]{mnras}

\usepackage{newtxtext,newtxmath}

\usepackage[T1]{fontenc}

\DeclareRobustCommand{\VAN}[3]{#2}
\let\VANthebibliography\thebibliography
\def\thebibliography{\DeclareRobustCommand{\VAN}[3]{##3}\VANthebibliography}

\usepackage{graphicx}	
\usepackage{amsmath}	
\usepackage{graphicx}
\usepackage{dcolumn}
\usepackage{bm}
\usepackage[dvipsnames]{xcolor}
\usepackage{xspace}
\usepackage{ulem}
\usepackage{lipsum}
\usepackage{multirow}
\usepackage{cancel}

\newcommand{\nv}{\hat{\bf n}}

\newcommand{\nmt}{{\tt NaMaster}\xspace}

\newcommand{\clin}{C_\ell^{\rm input}}

\title[Systematics mitigation for catalogue-based $C_\ell$s]{Systematics mitigation for catalogue-based angular power spectra}

\author[T.\ Cornish et al.]{
Thomas Cornish,$^{1,2}$\thanks{E-mail: thomas.cornish@physics.ox.ac.uk}
David Alonso,$^{1}$
Boris Leistedt$^{2}$
and Kevin Wolz$^{1}$
\\
$^1$Department of Physics, University of Oxford, Denys Wilkinson Building, Keble Road, Oxford OX1 3RH, United Kingdom\\
$^2$Department of Physics, Imperial College London, Blackett Laboratory, Prince Consort Road, London SW7 2AZ, United Kingdom
}

\date{Accepted XXX. Received YYY; in original form ZZZ}

\pubyear{2025}

\begin{document}
\label{firstpage}
\pagerange{\pageref{firstpage}--\pageref{lastpage}}
\maketitle

\begin{abstract}
Recent work has developed a formalism for computing angular power spectra directly from catalogues containing field values at discrete positions on the sky, thereby circumventing the need to create pixelised maps of the fields, as well as avoiding aliasing and finite-resolution effects. We adapt this formalism to incorporate template deprojection for mitigating systematic biases in the measured angular power spectra. We also introduce an alternative method of mitigating the `deprojection bias' {--} the loss of modes induced by deprojection {--} employing simple simulations to compute a transfer function. We find that this approach performs at least as well as existing methods, and is relatively insensitive to how well one can guess the true power spectrum of the observed field, except at the largest scales ($\ell \lesssim 3$). Additionally, we develop exact expressions for the bias introduced by deprojection in the shot-noise component, which further improves the accuracy of this approach. We test our formalism on simulated datasets, demonstrating its applicability both to discretely sampled fields, and to the special case of galaxy clustering, with the survey selection function defined in terms of a random catalogue or as a continuous sky map. After removing the bias in the shot noise and correcting for the remaining mode loss using a transfer function, our formalism produces unbiased measurements of the angular power spectrum in all scenarios tested here. Finally, we apply our formalism to real data and show it produces results consistent with the standard map-based pseudo-$C_\ell$ formalism. We implement our method in the public code \nmt. 
\end{abstract}




\section{Introduction}\label{sec:intro}
  The accurate measurement of projected fields on the sky has become a cornerstone of modern cosmology, with perhaps the most famous example being the cosmic microwave background (CMB), for which measurements of anisotropies in its temperature and polarisation enable constraints on models of inflation and the primordial density perturbations, as well as on the natures of neutrinos and dark energy \citep[e.g.][]{2020A&A...641A...6P, 2503.14452, 2025arXiv250620707C}. 
  Our knowledge of the large-scale structure (LSS) in the Universe is also heavily dependent on our ability to measure projected fields on the sky: tomographic measurements of galaxy clustering exploit the fact that galaxies are biased tracers of the underlying dark matter distribution while leveraging photometric redshifts as a means of estimating their radial positions, thereby enabling indirect measurements of the matter density field in `slices' across cosmic time without the need for spectroscopic observations, which are time-consuming and costly across large areas \citep[e.g.][]{2020JCAP...03..044N, 2105.13540, 2022PhRvD.105b3520A, 2023A&A...675A.202V, 2023PhRvD.108l3521S, 2025A&A...694A.259Y}. 
  Furthermore, the effects of weak gravitational lensing can be expressed in terms of projected fields, such as the lensing shear and convergence, both of which are sensitive probes of the intervening dark matter distribution \citep{2021A&A...646A.140H, 2022PhRvD.105b3520A, 2023PhRvD.108l3517M, 2025arXiv250319442S, 2025MNRAS.536.1586F}. These are only two examples of a myriad other projected probes of the LSS, which include other CMB secondary anisotropies, the Cosmic Infrared Background, etc. 
  \citep[e.g.][]{1999MNRAS.302..735S, 2017A&A...598A.115A, 10.3847/1538-4357/ab1f10, 2019MNRAS.489..401Z, 2019A&A...621A..32M, 2022JCAP...09..039C, 2022ApJS..258...36B, 2023MNRAS.520.1895J, 2024MNRAS.527.9581F, 2024MNRAS.528.2112S, 2024ApJ...962..112Q, 2024PhRvD.109f3530C, 2025PhRvD.111h3534G}.

  It is no surprise, then, that the angular power spectrum (denoted $C_\ell$), which quantifies the covariance between angular fluctuations in two fields as a function of angular multipole $\ell$ (and encodes all information in the case of Gaussian fields), has become an essential summary statistic in cosmology. Unbiased estimators of the angular power spectrum have been extensively explored, and optimal quadratic minimum variance (QMV) estimators have been developed which are capable of recovering angular power spectra with minimal loss of information \citep{astro-ph/9611174}. However, these estimators require the inversion of a pixel-space covariance matrix in order to optimally weight the data, an operation which can be computationally expensive when working with high-resolution maps. The prohibitive computational demand of these estimators often necessitates an alternative, an example of which can be found in the pseudo-$C_\ell$ (PCL) algorithm \citep{astro-ph/0105302,1809.09603}. The PCL approach simplifies things by assuming a diagonal data covariance matrix -- an assumption which holds if the data in question are spatially uncorrelated and/or dominated by white noise \citep{2013MNRAS.435.1857L}. This simplification significantly improves the speed of computation, making the PCL algorithm an often attractive approach to $C_\ell$ estimation.

  The characterisation and mitigation of systematic effects is becoming a problem of increasing relevance in cosmological surveys: as the hoards of data being collected continually increase in size, the statistical uncertainties have the potential to become comparable to those arising from systematics \citep[e.g.][]{10.1088/0004-637X/761/1/14,1212.4500,1211.1015,1309.2954,1410.0035,1311.2597,1709.08661}. 
  A plethora of techniques have been developed to tackle this issue by attempting to quantify the dependence of the desired field on maps of the survey properties or other contaminants thought to be contributing to the observed signal. These range from techniques interpretable through a linear regression framework \citep{2007.14499} such as mode deprojection \citep[also called template deprojection;][]{1992ApJ...398..169R,astro-ph/0403073,1509.08933,1609.03577}, template subtraction \citep{1105.2320,2012ApJ...761...14H}, and the iterative systematics decontamination (ISD) technique employed by the Dark Energy Survey \citep[DES;][]{1708.01536,2105.13540}, to more sophisticated methods employing machine learning tools such as artificial neural networks \citep{1907.11355} or self-organising maps \citep{2012.08467}.

  The approaches discussed so far typically involve data in the form of pixelised maps that quantify the value of a projected field at a given position on the sky. However, it is often the case in cosmology that the fields whose angular power spectra we would like to measure are sampled at discrete positions on the sky, corresponding to the positions of sources in a catalogue. The typical course of action to obtain power spectrum measurements in this case is to bin these discrete points into pixels and produce a map of the mean field value in each pixel. However, recent work by \citet{2312.12285,2407.21013} allows one to circumvent this step, through an update to the PCL formalism which facilitates the computation of the PCL directly from catalogue data. This approach naturally avoids any bias introduced by the process of pixelising discretely sampled observables, which would otherwise influence measurements of the power spectrum at scales comparable to or smaller than the pixel scale. 
  This extension to the PCL formalism was implemented in the publicly available code \nmt\footnote{\url{https://github.com/LSSTDESC/NaMaster}} \citep{1809.09603}, and its validity subsequently demonstrated on various discretely-sampled astrophysical fields, including cosmic shear, proper motions, dispersion measure of fast radio bursts (FRBs), and galaxy positions. However, these validation tests were performed using idealised mock data with no contamination due to systematics; indeed, the catalogue-based formalism as it is described in \cite{2407.21013} includes no means of mitigating systematic effects.

  The aim of this paper is thus to extend the catalogue-based PCL formalism of \cite{2407.21013} in order to incorporate systematics mitigation, and to implement this in \nmt. In particular, we adapt the template deprojection approach for applicability to catalogue data, as this is already implemented in \nmt for map-based fields. 
  
  The structure of the paper is as follows. In Section \ref{sec:meth} we outline the existing PCL formalism and its extension to catalogue-based data, before introducing our integration of template deprojection into the latter. In Section \ref{sec:res} we validate our approach using simulated data with systematic effects factored in. Section \ref{sec:conc} contains our conclusions.

\section{Methods}\label{sec:meth}
  \subsection{Overview of pseudo-\texorpdfstring{$C_\ell$}{Cl} methods}\label{ssec:meth.pcl}
    This section contains a brief review of the pseudo-$C_\ell$ power spectrum estimator, including the concept of template deprojection, and the techniques used to extract power spectra directly from catalogues. This provides the background needed to present the extension of deprojection techniques to catalogue-based spectra described in Section \ref{ssec:meth.deproj}. Readers are referred to \cite{astro-ph/0105302,1809.09603} for a more detailed account of these techniques.

    We will focus our discussion on the calculation of auto-correlations of a given field, since many of the new results presented in this paper are most relevant in this scenario, and all others can be trivially generalised to the case of cross-correlations.

    \subsubsection{Standard pseudo-\texorpdfstring{$C_\ell$}{Cl} estimator}\label{sssec:meth.pcl.pcl}
      The pseudo-$C_\ell$ (PCL) or MASTER algorithm for power spectrum estimation \cite{astro-ph/0105302,1809.09603} is a simplified version of the optimal QMV estimator. The optimal inverse-variance weighting of the data map, prescribed by the QMV estimator, is a computationally expensive operation, effectively requiring the inversion of a large pixel-space covariance matrix. In the PCL approach, this matrix is assumed to be diagonal, turning the inverse variance weighting step into a simple multiplication of two maps: the data map containing the field measurements, and the ``mask'', which plays the role of the local inverse variance of these measurements. Not only is this a significantly simpler and faster operation, but the statistical coupling between different angular multipoles induced by this weighting can be estimated analytically using efficient methods. This makes the PCL estimator easy to deploy on a large variety of datasets to extract fast power spectrum measurements. This is particularly beneficial on small scales, where the QMV estimator becomes prohibitively expensive.

      The steps involved in the PCL estimator can be summarised as follows (see \cite{1809.09603} for a more complete description of the algorithm)
      \begin{enumerate}
        \item Let ${\bf a}(\nv)$ be a spin-$s$ field defined on the sphere, and let $v(\nv)$ be its associated mask. Construct the masked field, given by the product ${\bf a}^v(\nv)\equiv v(\nv)\,{\bf a}(\nv)$. Note that we denote spin-$s$ quantities as a vector ${\bf a}(\nv)\equiv(a_1(\nv),a_2(\nv))$, where $a_1$ and $a_2$ are the two components of the field (e.g. the $Q$ and $U$ Stokes parameters for CMB polarisation maps, or the $\gamma_1$ and $\gamma_2$ components of the cosmic shear field).
        \item Compute the spherical harmonic transform of the masked field, ${\bf a}^v_{\ell m}$, ignoring the presence of a mask.
        \item Calculate the pseudo-$C_\ell$ of ${\bf a}^v_{\ell m}$, ${\rm PCL}_\ell({\bf a}^v,{\bf a}^v)$, defined as the na\"ive estimator of the field's power spectrum, averaging over all values of $m$ for a given $\ell$, ignoring the fact that ${\bf a}^v$ is in fact a masked field. In general, for two masked maps ${\bf a}^v$ and ${\bf b}^w$, the PCL\footnote{Note that we use ``PCL'' as shorthand for both the ``pseudo-$C_\ell$'' formalism and for the cut-sky estimator of the power spectrum defined here.} is defined as
        \begin{equation}
          {\rm PCL}_\ell({\bf a}^v,\,{\bf b}^w)\equiv\frac{1}{2\ell+1}\sum_{m=-\ell}^\ell {\bf a}^v_{\ell m}{\bf b}^{w\dagger}_{\ell m}\,.
        \end{equation}
        Since the presence of a sky mask leads to the statistical coupling of different angular multipoles, the PCL is also often called the ``coupled'' $C_\ell$.
        \item Calculate the ``mode-coupling matrix'' (MCM), the matrix encoding the statistical coupling between different $\ell$ modes induced by the sky mask, using the analytical expressions presented in e.g. \cite{1809.09603}. The MCM depends solely on the PCL of the mask. If binning the power spectrum into bandpowers, calculate also the binned MCM. Invert the MCM and multiply the PCL calculated in the previous step by it.
        \item In most cases, the measured field receives both signal and noise contributions: ${\bf a}={\bf s}_a+{\bf n}_a$, and more often than not we are interested in the power spectrum of the signal. Thus, an estimate of the noise power spectrum must be constructed and subtracted from the measured power spectrum. In certain cases (e.g. white noise) this may be done analytically. In other cases one must rely on noise simulations. Finally, noise bias may be avoided using only cross-correlations between different map ``splits'' (e.g. maps constructed from data taken during disjoint time periods), containing the same signal but different noise realisations \citep{2503.14452}. The specific case of uncorrelated noise from discretely-sampled sources is discussed in Section \ref{sssec:meth.pcl.cat}.
        \item The power spectrum must be corrected for finite-resolution effects. These may be caused by a finite telescope beam, or by the finite size of the sky pixels over which the map has been measured (assuming that the pixel values correspond to an average of the field value over the pixel area). This can be done exactly if a sufficiently accurate model of the smoothing kernel can be constructed \citep[e.g.][]{2024ApJS..273...26D}.
      \end{enumerate}
      It is worth noting that, under the usual interpretation of the QMV estimator, the mask $v(\nv)$ in step 1 above should in principle correspond to the local inverse variance of the measured field\footnote{We assume scalar masks, although this can be generalised to the case of fully anisotropic weighting -- see \cite{2410.07077}.}. However, the estimator remains unbiased regardless of what $v(\nv)$ is (as long as it is zero where the field is not measured), although its variance may become highly sub-optimal if $v(\nv)$ deviates significantly from this prescription.

    \subsubsection{Template deprojection}\label{sssec:meth.pcl.dpj}
      Often the map under study is affected by contamination from non-cosmological sources. Although the level of contamination is usually not known a priori, it is often possible to construct a comprehensive library of templates representing the potential sources of contamination. If the level of contamination is relatively mild, we can then construct a linear model of the form
      \begin{equation}\label{eq:lincont}
        {\bf a}^{\rm obs}={\bf a}^v(\nv)+A_p\,{\bf f}^p(\nv)\,,
      \end{equation}
      where ${\bf a}^{\rm obs}$ is the observed, contaminated field, ${\bf a}^v$ is the masked and uncontaminated field, ${\bf f}^p(\nv)$ are the different contaminant template maps, and $A_p$ are unknown amplitudes that control the level of contamination from each source. Note that we implicitly sum over the repeated index $p$ (and we will continue to do so throughout the paper).
      
      A self-consistent procedure to account for this contamination is then to calculate the power spectrum of ${\bf a}^{\rm obs}$ after marginalising over all possible values of $A_p$ \citep{1992ApJ...398..169R}. As shown in \cite{1609.03577}, in the context of the PCL framework, this can be achieved by calculating the best-fit value of $A_p$, and subtracting the resulting contamination from the original map. Mathematically, this is also equivalent to projecting the observed map onto the subspace that is orthogonal to all the contaminant templates, effectively ``de-projecting'' these templates from the data. The resulting deprojected map is thus
      \begin{equation}\label{eq:deproj_real}
        {\bf a}^c(\nv)={\bf a}^v(\nv)-{\bf f}^p(\nv)\,F_{pq}\left\langle {\bf f}^q,\,{\bf a}^v\right\rangle\,,
      \end{equation}
      where the matrix $F_{pq}$ is defined as the pseudo-inverse\footnote{The pseudo-inverse is used here to account for potential internal linear dependences in the library of contaminant templates.} of
      \begin{equation}\label{eq:F_inv}
        \left({\sf F}^{-1}\right)_{pq}\equiv \left\langle {\bf f}^p,\,{\bf f}^q\right\rangle\,.
      \end{equation}
      In these equations we have defined the map-level dot product as
      \begin{equation}\label{eq:dotprod}
        \left\langle{\bf a},\,{\bf b}\right\rangle\equiv\int d\nv\,{\bf a}^\dagger(\nv)\,{\bf b}(\nv)\,.
      \end{equation}
      Expanding the fields in terms of their harmonic coefficients, this dot product may also be written in harmonic space as
      \begin{equation}\label{eq:dotprod_harm}
        \left\langle{\bf a},\,{\bf b}\right\rangle=\sum_{\ell m}{\bf a}^\dagger_{\ell m}\,{\bf b}_{\ell m}=\sum_\ell(2\ell+1){\rm Tr}\,\left[{\rm PCL}_\ell({\bf a},{\bf b})\right]\,.
      \end{equation}
      Since this will be important for our discussion in Section \ref{ssec:meth.deproj}, note that the combination
      \begin{equation}\label{eq:a_coeff}
        \hat{A}_p\equiv F_{pq}\,\left\langle {\bf f}^q,\,{\bf a}^v\right\rangle
      \end{equation}
      is the best-fit value of the amplitude coefficient $A_p$ in Eq. \ref{eq:lincont} estimated from the data. Eq. \ref{eq:deproj_real} may thus be expressed directly in harmonic space:
      \begin{equation}\label{eq:deproj_harm}
        {\bf a}^c_{\ell m}={\bf a}^v_{\ell m}-\hat{A}_p\,{\bf f}^p_{\ell m}\,,
      \end{equation}
      with $\hat{A}_p$ calculated in real or harmonic space. Finally, note that all the template maps ${\bf f}^p$ entering the equations above are implicitly masked using $v(\nv)$ (the mask applied to the data).

      Since contaminant deprojection effectively removes modes from the map (any mode that lies in the subspace of contaminant templates), it causes a bias in the estimated power spectrum that must be corrected for \citep{1609.03577}. We will explore alternative approaches to correct for this bias in Section \ref{ssec:meth.dpbias}.

      As first introduced in Section 2.4.2 of \cite{1809.09603}, in the case of uncorrelated noise, it is possible to calculate the correction to the noise power spectrum caused by contaminant deprojection exactly, even if the noise is inhomogeneous. The additional contribution to the noise PCL is:
      \begin{align}\nonumber
        \Delta {\sf N}_\ell = 
        &-2F_{pq}\,{\rm PCL}_\ell({\bf f}^p,v^2\sigma_N^2{\bf f}^q)\\\label{eq:noise_deproj}
        &+F_{pq}F_{rs}\left[\int d\nv\,v^2\sigma_N^2{\bf f}^{q\dagger}{\bf f}^r\right]\,{\rm PCL}_\ell({\bf f}^p,{\bf f}^s)\,,
      \end{align}
      where $\sigma_N^2(\nv)$ is a map of the local field variance in a unit solid angle.

    \subsubsection{Catalogue-based pseudo-\texorpdfstring{$C_\ell$}{Cl}s}\label{sssec:meth.pcl.cat}
      We are often interested in fields that are not mapped over continuous areas, but instead are recorded at the positions of discrete sources. A prime example of this is cosmic shear, where the lensing shear field is effectively measured at the positions of background sources. The power spectrum of such fields can still be calculated using the standard PCL approach, generating continuous maps by binning the sources into pixels, and using the density of these sources as a proxy for the sky mask. Some subtleties about this approach can be found in \cite{2010.09717}.

      Instead, a more accurate treatment of discretely-sampled fields has been introduced recently \citep{2312.12285,2407.21013}, enabled by new algorithms to estimate the spherical harmonic transforms (SHTs) of irregularly-sampled fields \citep{2304.10431}. In this case, the field and its mask can be expressed as a sum of delta functions centred at the source coordinates:
      \begin{align}\label{eq:cat_field}
        v(\nv)=\sum_{i=1}^{N_d} w_i\delta^D(\nv,\nv_i)\,,\hspace{12pt}
        a^v(\nv)=\sum_{i=1}^{N_d} w_i\,a_i\,\delta^D(\nv,\nv_i)\,,
      \end{align}
      where the index $i$ runs over all sources in the sample, $w_i$ is the weight of the $i$-th source, $a_i$ is the value of field measured at it, $\nv_i$ is its position in the sky, and $\delta^D$ is the Dirac delta function on the sphere. For simplicity we consider only a scalar field here, and the generalisation to spin-$s$ can be found in \cite{2407.21013}. The spherical harmonic coefficients of $v$ and $a^v$ are then simply
      \begin{align}
        v_{\ell m}=\sum_{i=1}^{N_d}w_i\,Y_{\ell m,i}^*\,,\hspace{12pt}
        a^v_{\ell m}=\sum_{i=1}^{N_d}w_i\,a_i\,Y_{\ell m,i}^*\,,
      \end{align}
      which can be calculated using the irregular SHT methods implemented in e.g. \cite{2304.10431}. Once computed, these harmonic coefficients can be used to calculate the PCL of the field and the associated MCM, as described in Section \ref{sssec:meth.pcl.pcl}.

      The most immediately obvious advantage of the catalogue-based approach is the absence of finite pixel resolution effects, allowing the resulting estimator to recover power spectra up to arbitrarily small scales (as long as the corresponding harmonic coefficients can be calculated and stored). In addition to this, \cite{2407.21013} showed that the catalogue-based approach naturally allows for a more numerically stable treatment of uncorrelated noise bias. In particular, it was shown that all contributions from any uncorrelated noise sources can be removed exactly by subtracting the zero-lag contributions to the PCLs of the masked field and of the mask itself prior to estimating the mode-coupling matrix (i.e. the contributions to the PCLs coming from the product of $v$ and $a^v$ with themselves evaluated at the same source). These are given by
      \begin{equation}\label{eq:cat_noise}
        \tilde{N}^v\equiv \frac{1}{4\pi}\sum_{i=1}^{N_d}w_i^2\,,\hspace{12pt}
        \tilde{N}^a\equiv \frac{1}{4\pi}\sum_{i=1}^{N_d}w_i^2a^2_i\,.
      \end{equation}
      Doing so has the added advantage that, by removing the asymptotic value of the mask PCL at high $\ell$ (given by $\tilde{N}^v$), the associated MCM is significantly more numerically stable, and can be safely inverted.

      Another application of the catalogue-based approach is the case of galaxy clustering, in which the density of sources is itself the field of interest. In the simplest scenario, a continuous map with the expected mean density of sources $\bar{n}(\nv)$ can be constructed. The overdensity field is then given by
      \begin{equation}
        a(\nv)\equiv\delta_g(\nv)=\frac{1}{\bar{n}(\nv)}\left[\sum_{i=1}^{N_d} w_i\delta^D(\nv,\nv_i)-\bar{n}(\nv)\right]\,.
      \end{equation}
      Assuming the mean density is itself used as a mask for this field (since $\bar{n}(\nv)$ is indeed proportional to the inverse variance of $\delta_g$ in the Poisson limit), the masked field is simply
      \begin{equation}\label{eq:av_gc}
        a^v(\nv)=\sum_{i=1}^{N_d}w_i\delta^D(\nv,\nv_i)-\bar{n}(\nv)\,.
      \end{equation}
      Since in this case the mask is continuous, its shot noise component is simply $\tilde{N}^v=0$, while the field noise is $\tilde{N}^a=\sum_{i=1}^{N_d} w_i^2/(4\pi)$. 
      
      Often, however, the mean source density is not provided as a continuous map, but as a catalogue of ``random'' objects (or simply ``randoms''). Catalogues of randoms have long been a critical component of configuration-space galaxy clustering analyses \citep[e.g.][]{1969PASJ...21..221T, 1993ApJ...412...64L}. These are point processes distributed according to the survey selection function and without any intrinsic clustering. They are typically generated by uniformly sampling across the survey footprint to create a catalogue of positions with a density significantly higher than that of the data, such that correlations between the randoms and observed galaxies can be used to mitigate systematics caused by masking or edge effects. It is worth noting that recent efforts have also been made to construct catalogues of `organised randoms' which additionally account for spatial correlations induced by more complex survey properties such as inhomegeneous depth or seeing \citep{2012.08467, 2025A&A...694A.259Y}. However, since the aim of this work is to employ template deprojection to mitigate these more complex effects, we adopt the former approach to generating randoms when applying our formalism to simulations (Section \ref{ssec:res.gcval}) and observations (Section \ref{ssec:res.data}).
      
      If randoms are used in place of a continuous map, the mean density map (and hence the mask) entering the equations above is
      \begin{equation}\label{eq:n_rand}
        \bar{n}(\nv)=\alpha\,\sum_{i=1}^{N_r} w^r_i\,\delta^D(\nv,\nv_i)\,,
      \end{equation}
      where $w_i^r$ is the weight of the $i$-th random object, and $\alpha\equiv\sum_{i=1}^{N_d}w_i/\sum_{i=1}^{N_r}w_i^r$ corrects for the different number of sources in the data and random catalogues. To simplify the notation, in what follows, we will omit the factor $\alpha$, and absorb it into the random weights\footnote{In other words, we will implicitly assume that the random weights are normalised such that $\sum_{i=1}^{N_r}w_i^r=\sum_{i=1}^{N_d}w_i$.}. In this case the mask has a shot noise component, and the randoms also contribute to the field noise:
      \begin{equation}
        \tilde{N}^w=\frac{1}{4\pi}\sum_{i=1}^{N_r}(w_i^r)^2\,,\hspace{12pt}
        \tilde{N}^a=\frac{1}{4\pi}\sum_{i=1}^{N_d}\,w_i^2+\tilde{N}^w\,.
      \end{equation}

      In summary, the treatment of catalogue-based fields is straightforward to extend to the case of galaxy clustering, by simply replacing all field values with 1, and including the contribution from random objects in the calculation of the field spherical harmonic coefficients and in the shot-noise components of field and mask (if randoms are used).

  \subsection{Template deprojection for catalogue-based \texorpdfstring{$C_\ell$}{Cl}s}\label{ssec:meth.deproj}
    We now present the extension of the template deprojection technique, outlined in Section \ref{sssec:meth.pcl.dpj} for continuous fields, to the case of fields sampled at the positions of discrete sources. Specifically, our aim is to extend Eqs. \ref{eq:deproj_harm} and \ref{eq:noise_deproj}, describing the harmonic coefficients of the contaminant-deprojected field and the modification to the white noise level brought about by deprojection, respectively. We will do so for both sampled fields and for galaxy clustering.

    \subsubsection{Sampled fields}\label{sssec:meth.deproj.sampled}
      Consider first a discretely-sampled field $a^v(\nv)$ as given in Eq. \ref{eq:cat_field}. The masked contaminant templates in this case are
      \begin{equation}
        \tilde{f}^p(\nv)=\sum_i w_i\,f^p_i\,\delta^D(\nv,\nv_i)\,,
      \end{equation}
      where $f_i^p$ is the value of the template at the position of the $i$-th source in the catalogue. Substituting this in Eq. \ref{eq:deproj_real}, we find that the deprojected field is simply
      
      \begin{equation}
        a^c(\nv)=\sum_iw_i\,a^c_i\,\delta^D(\nv,\nv_i)\,,
      \end{equation}
      where we have defined the deprojected sampled field values $a^c_i$ are
      \begin{equation}\label{eq:deproj_sampled}
        a^c_i\equiv a_i-\hat{A}_p\,f_i^p\,,
      \end{equation}
      and the estimated linear amplitude is
      \begin{align}
        \hat{A}_p\equiv\tilde{F}_{pq}\sum_j w_j^2 f_j^qa_j\,,\hspace{12pt}
        \left(\tilde{\sf F}^{-1}\right)_{pq}\equiv \sum_jw_j^2f_j^pf_j^q\,.
      \end{align}
      Thus, we see that deprojection simply affects the values of the field sampled at the source positions, and it only depends on the values of the contaminant templates sampled at the same positions. The dot product introduced in Eq. \ref{eq:dotprod} is replaced by a sum over all sources of the product of two weighted fields. Deprojection can therefore be achieved easily, without requiring any contaminant maps explicitly, and simply modifying the sampled field values as above. Likewise, the white noise amplitude, accounting for deprojection, can be calculated simply using Eq. \ref{eq:cat_noise}, replacing $a_i$ with $a_i^c$ as given above.

      The white-noise power spectrum receives a further scale-dependent correction from deprojection due to mode loss. In the case of continuous fields, this is given by Eq. \ref{eq:noise_deproj}. To calculate this correction in the case of a discretely-sampled field, we start by writing the PCL of $a^c$:
      \begin{equation}
        {\rm PCL}_\ell(a^c,a^c)=\sum_{ij}(w_i\,w_j\,a_i^c\,a_j^c)\,\sum_{m=-\ell}^\ell\frac{\left|Y^*_{\ell m,i}Y_{\ell m,j}\right|}{2\ell+1}\,.
      \end{equation}
      We then expand $a^c_i$ in terms of $a_i$ and the templates as in Eq. \ref{eq:deproj_sampled}, and then take the ensemble average of the result, assuming $a_i$ to consist solely of uncorrelated noise: $\langle a_ia_j\rangle=\sigma_i^2\,\delta_{ij}$, where $\sigma_i^2$ is the noise variance of the field at the $i$-th discrete source. Writing the result as
      \begin{equation}
        \left\langle{\rm PCL}_\ell(a^c,a^c)\right\rangle=\tilde{N}^a+\Delta N_\ell^a\,,
      \end{equation}
      the correction is
      \begin{align}\label{eq:noise_deproj_sampled}
        \Delta N^a_\ell=&-2\tilde{F}_{pq}\,{\rm PCL}_\ell(\tilde{f}^p,\tilde{g}^q)\\\nonumber
        &+\tilde{F}_{pq}\tilde{F}_{rs}\,\left[\sum_jw_j^2\sigma^2_j\,(w_jf^q_j)\,(w_jf^s_j)\right]{\rm PCL}_\ell(\tilde{f}^p,\tilde{f}^r)\,,
      \end{align}
      where
      \begin{equation}
        \tilde{f}^p_{\ell m}\equiv\sum_iw_if_i^p\,Y^*_{\ell m,i}\,,
      \end{equation}
      and where we have defined the noise-weighted field
      \begin{equation}
        \tilde{g}^p_{\ell m}\equiv\sum_iw_i^2\sigma_i^2\,(w_if_i^p)\,Y^*_{\ell m,i}\,.
      \end{equation}

      As before, this contribution to the noise bias due to deprojection depends only on the values of the contaminant templates evaluated at the source positions, as well as an estimate of the noise variance at each source $\sigma_i^2$. For noise-dominated fields, this may be estimated directly as $\sigma_i^2\sim a_i^2$, but care must be exercised when this is not the case. For modest numbers of contaminant templates, this white noise correction is relatively small, and may be neglected. However, as we will see in Section \ref{ssec:res.sampval}, when many ($O(100)$) templates are deprojected, the correction may become significant, particularly on large angular scales.

    \subsubsection{Galaxy clustering}\label{sssec:meth.deproj.gc}
      We turn now to the case of galaxy clustering. We will start with the more complex scenario in which the mask is provided as a discrete set of random points, and then turn to the simpler case of a continuous mask at the end of this section.
      
      Two characteristics of this type of data separate it from the standard sampled fields in a way that impacts how contaminant deprojection must be carried out: first, the source positions and weights do not constitute the field's mask, the random catalogue does instead. Secondly, this discretely-sampled mask features as an additive contribution to the masked field itself, quantifying the mean density subtracted from the data (the second term in Eq. \ref{eq:av_gc}). Note that although this latter choice is indeed the most common one when analysing galaxy data, one could also use two different random catalogues -- one to construct the subtracted mean density, and a different one to define the mask -- and the resulting estimator for the galaxy overdensity would still be unbiased\footnote{Assuming that both random catalogues represent the same underlying mean density field.}. Nevertheless, we will adopt the standard approach of using a single random catalogue for both purposes here.

      This immediately leads to a problem if we simply apply Eq. \ref{eq:deproj_real} to the field defined in Eq. \ref{eq:av_gc}, with $\bar{n}$ given by Eq. \ref{eq:n_rand} and, hence, with the masked templates given by\footnote{Recall that we assume the random weights to be normalised so that $\alpha=1$.}
      \begin{equation}\nonumber
        \tilde{f}^p(\nv)=\sum_{i=1}^{N_r}w_i^r f_i^p\,\delta^D(\nv,\nv_i)\,,\hspace{12pt}
        \tilde{f}^p_{\ell m}=\sum_{i=1}^{N_r}w_i^r f_i^p\,Y^*_{\ell m,i}\,.
      \end{equation}
      Substituting this in the expression for the estimated linear coefficients $\hat{A}_p$ (Eq. \ref{eq:a_coeff}) we find:
      \begin{align}\nonumber
        \hat{A}_p&=F_{pq}\int d\nv\,a^v(\nv)\,\tilde{f}^q(\nv)\\\nonumber
               &=F_{pq}\left(\sum_{i=1}^{N_d}\sum_{j=1}^{N_r}w_iw_j^rf_j^q\int d\nv \delta^D(\nv_i,\nv)\delta^D(\nv_j,\nv)\right.\\\nonumber
               &\hspace{39pt}\left.-\sum_{i,j=1}^{N_r}w_i^rw_j^rf_j^q\int d\nv \delta^D(\nv_i,\nv)\delta^D(\nv_j,\nv)\right)\\\nonumber
               &\propto -F_{pq} \sum_{i=1}^{N_r}(w^r_i)^2f^p_i\,,
      \end{align}
      since we assume that no two objects from either the source or random catalogues occupy the exact same position. Likewise, the elements of the $F_{pq}$ matrix are
      \begin{equation}
        \left({\sf F}\right)^{-1}_{pq}\propto \sum_{i=1}^{N_r}(w_i^r)^2f_i^pf_i^q\,.
      \end{equation}
      This is problematic because, first, the estimated $\hat{A}_p$ does not depend on the data catalogue, only on the randoms, and hence it does not capture any real correlations between the galaxy overdensity and the contaminants. Secondly, had we chosen to use two different catalogues for the subtracted mean density and the mask, the quantities above would both be exactly zero.

      The cause of this problem lies in the key assumption underlying the pseudo-$C_\ell$ power spectrum estimator: the covariance matrices of all fields involved are diagonal in real space, and the inverse diagonals are represented by the field masks (in this case, the random catalogue positions). We know that, in practice, this is not the case: LSS fields exhibit correlations over a range of real-space separations. Furthermore, we actually expect the observed galaxy overdensity to be affected by systematic contamination (and hence to correlate with the contaminant templates) on large scales. We thus propose to use these large-scale correlations to redefine the linear deprojection coefficients. In particular, we can use the harmonic-space expression for the dot product of two fields (Eq. \ref{eq:dotprod_harm}), truncating it at a given maximum multipole $\ell_{\rm max}$, which characterises the scale over which we want to capture correlations between data and contaminants. The linear deprojection coefficients, in the case of galaxy clustering, are therefore calculated via
      \begin{align}\label{eq:a_coeff_gc}
        &\hat{A}_p\equiv \tilde{F}_{pq}\,\sum_{\ell=0}^{\ell_{\rm max}}(2\ell+1)\left[{\rm PCL}_\ell(a^v,\tilde{f}^q)+\sum_{i=1}^{N_r}\frac{(w_i^r)^2\,f_i^q}{4\pi}\right]\,,\\\nonumber
        &\left(\tilde{\sf F}\right)^{-1}_{pq}=\sum_{\ell=0}^{\ell_{\rm max}}(2\ell+1)\left[{\rm PCL}_\ell(\tilde{f}^p,\tilde{f}^q)-\sum_{i=1}^{N_r}\frac{(w_i^r)^2\,f_i^pf_i^q}{4\pi}\right]\,.          
      \end{align}
      Note that we have subtracted the shot-noise contribution to the pseudo-$C_\ell$s entering these equations, caused by coincident pairs of random points in the subtracted mean density and the mask. This contribution is completely accidental (e.g. it would be zero if we had decided to use two different random catalogues for both purposes), and must therefore be discarded.

      This generalisation of the standard contaminant deprojection approach is perfectly legitimate from a mathematical point of view: it is equivalent to a standard deprojection operation in harmonic space in which some data points (multipoles with $\ell,|m|>\ell_{\rm max}$) are assigned zero weight (or, equivalently, a very large variance). 

      Having defined the deprojection procedure, we must now derive the correction to the shot noise power spectrum caused by deprojection. We can do so following the same steps outlined in the previous section for sampled fields, with two caveats:
      \begin{itemize}
        \item The noise variance in this case is sourced by Poisson noise in the data and random catalogues. Furthermore, if different random catalogues are used to quantify the subtracted mean density and to define the mask, only the randoms in the former (i.e. those entering $a^v_{\ell m}$) contribute to this variance, and not those in the latter (i.e. those entering $\tilde{f}_{\ell m}$).
        \item We must take care to consistently apply the truncation scale $\ell_{\rm max}$ used to find the deprojection coefficients.
      \end{itemize}
      The resulting expression for the shot noise deprojection bias is
      \begin{align}\nonumber
        \Delta N_\ell=
        &-2\tilde{F}_{pq}\,{\rm PCL}_\ell(\tilde{f}^p,\tilde{g}^q)\\\nonumber
        &+\tilde{F}_{pq}\tilde{F}_{rs}\left[\sum_{i=1}^{N_d}w_i^2\tilde{f}_{Fi}^q\tilde{f}_{Fi}^s\right.\\\label{eq:deltaN_gc}
        &\hspace{43pt}\left.+\sum_{i=1}^{N_r}(w_i^r)^2\tilde{f}_{Fi}^q\tilde{f}_{Fi}^s\right]\,{\rm PCL}_\ell(\tilde{f}^p,\tilde{f}^q)\,,
      \end{align}
      where we have defined the filtered-weighted templates $\tilde{f}^p_{F}$ as a truncated inverse spherical harmonic transform of the masked templates
      \begin{equation}
        \tilde{f}^p_{Fi}\equiv \sum_{\ell=0}^{\ell_{\rm max}}\sum_{m=-\ell}^\ell\tilde{f}^p_{\ell m}\,Y_{\ell m,i}\,,
      \end{equation}
      and where
      \begin{equation}\label{eq:glm_gc}
        \tilde{g}^p_{\ell m}\equiv\sum_{i=1}^{N_d}w_i^2\tilde{f}^p_{Fi}Y^*_{\ell m,i}+\sum_{i=1}^{N_r}(w_i^r)^2\tilde{f}^p_{Fi}Y^*_{\ell m,i}\,.
      \end{equation}

      Note that, as in the case of sampled fields, we only require the values of the contaminant templates at the discrete positions of the points defining the field's mask (i.e. the random catalogue), and not explicitly in the form of maps.
      
      Now, in the simpler case in which the mask (and hence the mean number density) is provided as a continuous map, deprojection can be carried out following the steps outlined above, with the following simplifications:
      \begin{itemize}
        \item The templates are also provided as continuous maps, and the masked templates are simply $\tilde{f}^p(\nv)=\bar{n}(\nv)\,f^p(\nv)$, where $\bar{n}(\nv)$ is the mean number density map, which also serves as our mask.
        \item No shot-noise contributions must be subtracted when estimating the deprojection coefficients (i.e. the rightmost terms in both expressions in Eq. \ref{eq:a_coeff_gc}).
        \item Only the data weights contribute to the shot noise variance in Eqs. \ref{eq:deltaN_gc} and \ref{eq:glm_gc}.
      \end{itemize}


  \subsection{Correcting for mode loss}\label{ssec:meth.dpbias}
    The process of template deprojection indiscriminately removes all power from modes proportional to any of the deprojected templates, introducing a bias in the final power spectrum estimate. As discussed in \cite{1609.03577}, this {\it deprojection bias} can be estimated analytically and subsequently removed: for a scalar field $a$ with corresponding mask $v$ and contaminants $f^p$, it is given by
    \begin{align}\nonumber
      \Delta C_\ell=&-2F_{pq}\,{\rm PCL}_\ell\left(\tilde{f}^q_{\ell m},f^p_{\ell m}\right)\\\label{eq:dpbias}
        &+F_{pq}F_{rs}\left[\int d\nv\,f^{q\dag}(\nv)\tilde{f}^r(\nv)\right]{\rm PCL}_\ell\left(f^p,f^s\right),
    \end{align}
    where $F_{pq}$ is as defined via Eq.~\ref{eq:F_inv}, and
    \begin{equation}\label{eq:dp.f.aux}
      \tilde{f}^p_{\ell m}\equiv{\cal S}\left[v(\nv){\cal S}^{-1}\left[{C}^{\rm true}_{\ell^\prime}{\cal S}\left[v\,f^p\right]^{s_b}_{\ell^\prime m^\prime}\right]^{s_a}_{\nv}\right]^{s_a}_{\ell m}.
    \end{equation}
    with ${\cal S}$ denoting a spherical harmonic transform and ${\cal S}^{-1}$ its inverse\footnote{Defined as ${\cal S}\left[ a(\nv) \right]_{\ell m} \equiv \int d\nv Y_{\ell m}^\ast(\nv) a(\nv)$ and ${\cal S}^{-1} \left[ a_{\ell m} \right]_{\nv} \equiv \sum_{\ell m} Y_{\ell m}(\nv) a_{\ell m}$ for a scalar field $a(\nv)$.}. 
    Note that we are again only considering autocorrelations of a single scalar field here; a generalisation to cross-correlations between two fields with arbitrary spin can be found in \cite{1809.09603}.

    This method of correcting for the loss of modes induced by deprojection has two drawbacks. Firstly, it requires some knowledge of the true angular power spectrum $C^{\rm true}_\ell$, for which a reliable estimate is not always achievable. Secondly, while the above prescription works for map-based fields, an equivalent procedure for catalogue-based data is non-trivial to implement, owing to the inverse spherical harmonic transforms involved.

    We therefore explore in this work an alternative approach, in which we use simulations to estimate a \textit{transfer function} by which one can divide the measured power spectra to correct for mode loss following template deprojection. The process involves randomly generating many Gaussian realisations of some angular power spectrum $\clin$, and applying to each realisation the same mask as is being used for the data in question. For each realisation, the angular power spectrum is then estimated both before and after deprojecting the set of contaminants believed to be influencing the data; we will denote these power spectra as $C_\ell^{\rm before}$ and $C_\ell^{\rm after}$, respectively. 
    The transfer function $T_\ell$ is then defined as
    \begin{equation}\label{eq:transfer_func}
      T_\ell = \frac{\langle C_\ell^{\rm after} \rangle_{\rm sim}}{\langle C_\ell^{\rm before} \rangle_{\rm sim}}\,,
    \end{equation}
    where $\langle \cdot \rangle_{\rm sim}$ denotes the average across Gaussian realisations. Dividing the power spectrum measured from the data by $T_\ell$ should then correct it for mode loss (see e.g. \citealt{1509.08933}, in which this approach is used to validate analytical expressions for the deprojection bias). 

    In Section \ref{ssec:res.db_vs_tf} we test this approach and verify it as a viable alternative to calculating the deprojection bias, which can in turn be applied to power spectra measured from catalogue-based data. 

\section{Results}\label{sec:res}
  We now validate the catalogue-based adaptation of template deprojection described above using both simulated and real data. Sections \ref{ssec:res.sampval} and \ref{ssec:res.gcval} do this for the cases of sampled fields and galaxy clustering, respectively.  Next, in Section \ref{ssec:res.db_vs_tf} we examine the relative performances of the mode loss correction methods described in Section \ref{ssec:meth.dpbias}. 
  Finally in Section \ref{ssec:res.data} we apply this approach to real data from the {\it Gaia-unWISE} quasar catalogue \citep[{\it Quaia};][]{2306.17749}.

  \subsection{Contaminant deprojection in sampled fields}\label{ssec:res.sampval}
    \begin{figure}
      \centering
      \includegraphics[width=0.49\textwidth]{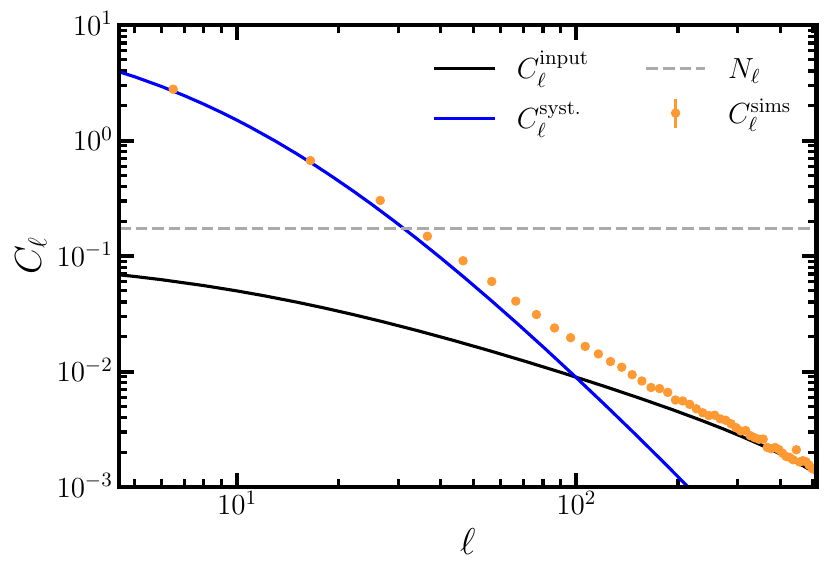}
      \caption{Input power spectrum used to validate the approach to contaminant deprojection in the case of sampled fields (black line). The blue line shows the power spectrum of the contaminant maps used, while the horizontal dashed line shows the simulated white noise component. The orange points show the mean power spectrum measured from all contaminated simulations.}
      \label{fig:sampled_input}
    \end{figure}

    \begin{figure}
      \centering
      \includegraphics[width=0.49\textwidth]{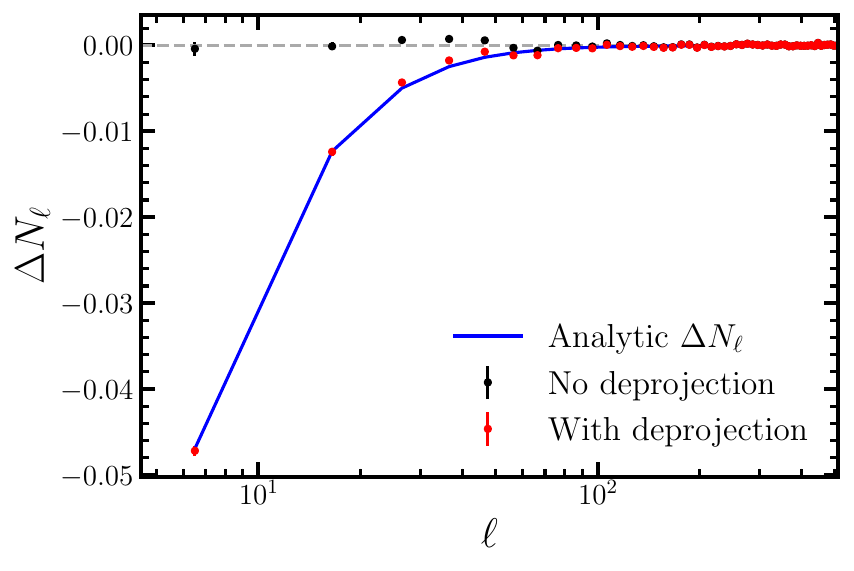}
      \caption{Power spectrum of noise-only simulations with and without contaminant deprojection (black and red points, respectively). The loss of power due to deprojection leads to a negative power spectrum (note that the catalogue-based estimator automatically removes the white noise component), which is accurately predicted by Eq. \ref{eq:noise_deproj_sampled} (blue line).}
      \label{fig:sampled_noideproj}
    \end{figure}

    We start by validating the formalism described in the previous section in the case of sampled fields. To do so, we generate a number of simulated catalogue-based observations including the presence of sky contamination, and show that we are able to recover unbiased power spectra by applying deprojection techniques and accounting for the impact of deprojection bias in both the signal and uncorrelated noise components.

    Specifically, in our simulations we use the positions of sources in the {\sl Quaia} catalogue with spectro-photometric redshifts in the range $z_p<1.47$, corresponding to a sample of $644786$ quasar candidates. This ensures that the simulated data have a realistic source distribution, including large-scale sky coverage and small-scale clustering. Using a {\tt HEALPix}\footnote{\url{http://healpix.sf.net/}}\xspace \citep{2005ApJ...622..759G,Zonca2019} pixelisation scheme with a resolution parameter of $N_{\rm side} = 256$ (corresponding to a pixel area of ${\sim}0.05$~deg$^2$), we then generate signal maps as Gaussian realisations with an angular power spectrum of the form
    \begin{equation}
      C_\ell^{\rm input}=\frac{1}{\ell+10}\,,
    \end{equation}
    and sample them at the positions of the {\sl Quaia} sources. We generate $1000$ such realisations. Likewise, we generate $100$ Gaussian maps from a steeper power spectrum of the form
    \begin{equation}
      C_\ell^{\rm syst}=\frac{A_{\rm syst}}{(\ell+10)^3}\,,
    \end{equation}
    which we sample at the same source positions and use as contaminant templates to deproject. We then construct a specific linear combination of these contaminant templates to contaminate the signal assigned to each source. For simplicity, we use the direct sum of all templates, scaling the proportionality constant $A_{\rm syst}$ to ensure that the total variance due to systematic contamination in each source is $30\%$ of the signal. Finally, we generate uncorrelated Gaussian numbers for each source to serve as a noise component, with a standard deviation equal to 10 times that of the signal. Figure \ref{fig:sampled_input} shows the input and contaminant power spectra, as well as the expected noise level, together with the mean power spectrum of all simulations. We note that the amplitudes of the contaminant and noise components have been chosen to exacerbate the impact of systematic deprojection on the simulations: the measurements are systematics-dominated on scales $\ell\simeq100$. Furthermore, the data are noise dominated on a source-by-source basis, as well as on all scales. This, together with the large number of contaminant templates being deprojected, increases the amplitude of the correction to the noise bias due to deprojection ($\Delta N_\ell^a$ in Eq. \ref{eq:noise_deproj_sampled}).
    \begin{figure}
      \centering
      \includegraphics[width=0.49\textwidth]{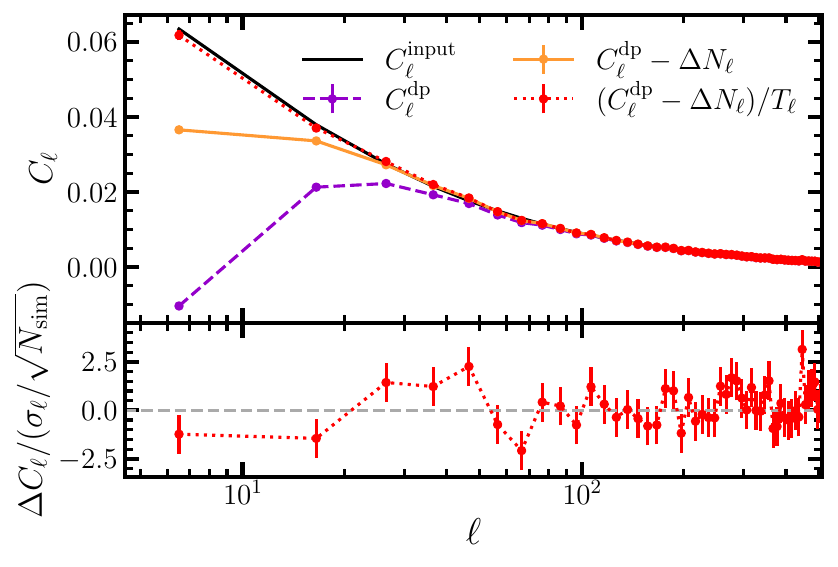}
      \caption{Mean and standard deviation of the power spectra from contaminated simulations after contaminant deprojection (purple points). The orange points show the result of accounting for the impact of deprojection in the white noise component (via Eq. \ref{eq:noise_deproj_sampled}), while the red points show the power spectra after accounting for deprojection bias in both the noise and signal components (the latter via the transfer function approach). The result is an unbiased estimate of the input power spectrum (black line). The bottom panel shows the residuals with respect to the input spectrum normalised by the error in the mean over all simulations.}
      \label{fig:sampled_validation}
    \end{figure}

    To first validate our analytical expression used to calculate $\Delta N_\ell^a$, we study the power spectrum of noise-only simulations, in which we only include the contribution from uncorrelated noise to the field value for each source. Fig. \ref{fig:sampled_noideproj} shows the mean power spectrum of all such simulations before and after deprojecting all 100 contaminant templates (black and red points, respectively). We remind the reader that the white noise component is avoided by the catalogue-based estimator by construction. The loss of modes due to deprojection leads to a negative power spectrum at low $\ell$, which is accurately recovered by our analytical expression (blue line).

    Finally, Fig. \ref{fig:sampled_validation} shows the result of our validation of the formalism outlined here. The purple points show the average power spectrum for contaminated simulations after contaminant deprojection, before accounting for any sources of deprojection bias. On large scales, the power spectrum shows a very significant lack of power, even becoming negative at the lowest $\ell$s. Much of this power is recovered by correcting for the white noise deprojection bias calculated via Eq. \ref{eq:noise_deproj_sampled}, as shown by the orange points. The remaining power loss can be recovered by correcting for the deprojection bias of the signal component. As described in Section \ref{ssec:meth.dpbias}, this may be done analytically if the true signal power spectrum is known, or through a simulation-based transfer function approach. The red points show the result of the latter method\footnote{It is worth noting that the true signal power spectrum is used here to generate the simulations required for deriving the transfer function (see Section \ref{ssec:meth.dpbias}). However, as we will show in Section \ref{ssec:res.db_vs_tf}, this approach is relatively insensitive to this choice of power spectrum.}, which is able to recover an unbiased power spectrum at all scales.

  \subsection{Contaminant deprojection in galaxy clustering}\label{ssec:res.gcval}
    Here we validate the galaxy clustering formalism described in Section \ref{sssec:meth.deproj.gc} by again using simulated catalogue-based observations, with contributions from sky contaminants imposed on the distribution of the sources. We assess the two possible treatments of the mask: using the positions and weights of randoms to define it, and providing it as a continuous map. We will hereafter refer to these as the randoms-based and mask-based approaches, respectively.

    For each simulation, we begin by generating a Gaussian realisation of a map, $\delta_g^{\rm in}$, from some input angular power spectrum, $\clin$. For this test, we use the best-fit angular power spectrum for the {\sl Quaia} `low-$z$' ($0 \leq z < 1.47$) sample, described in \cite{2410.24134,2504.20992}, as the input for our simulations. A second map, $\delta_g^{\rm cont}$, is also generated from $\clin$ and contaminated with templates $f^p$ using known linear coefficients $A_p$ such that the value in pixel $i$ is given by
    \begin{equation}\label{eq:dg_cont}
      \delta_{g,i}^{\rm cont} = \delta_{g,i}^{\rm in} + \sum_{p = 1}^{N_{\rm syst}} A_p f_i^p\,.
    \end{equation}
    We use the same 10 templates as used in \cite{2410.24134,2504.20992}, which the authors deproject from {\sl Quaia} overdensity maps using the standard formalism as outlined in Section\ \ref{ssec:meth.deproj}. 
    The use of these templates here is motivated by the desire to test our catalogue-based formalism in a realistic scenario for galaxy clustering. 
    Details of these templates can be found in Appendix \ref{sec:appx.templates}, including Mollweide projections of each (Fig.\ \ref{fig:templates}).
    
    We then use a mask defining the {\sl Quaia} survey geometry to convert each of these maps into a probability map, $P$, whose value in a pixel $i$ is given by: 
    \begin{equation}\label{eq:pmap}
      P^X_i = (1 + \delta_{g,i}^X) w^i\,,
    \end{equation}
    where $w$ is a map describing the survey mask, and $X \in \{{\rm in}, {\rm cont}\}$. As in Section \ref{ssec:res.sampval}, all maps are created using a {\tt HEALPix}\xspace pixelisation scheme with a resolution parameter of $N_{\rm side} = 256$. These maps are then normalised such that $\max(P^X) = 1$, and subsequently Poisson-sampled to create realisations of the catalogue with and without contamination from systematics. We note that it is possible, when generating these probability maps, for a pixel to end up with a value less than zero, and cannot therefore be Poisson-sampled. To test whether this is likely to be an issue, we use the prescription outlined above to generate two sets of $10^5$ probability maps, one with contamination and one without, and count how many pixels in each have values below zero. We find that there are no such pixels in any of the maps, and thus conclude that they are unlikely to contribute significantly to our analysis. 

    As highlighted in Section \ref{sssec:meth.deproj.gc}, it is necessary in the case of galaxy clustering systematics to truncate at some maximum multipole $\ell_{\rm max}$ the harmonic-space expression for the dot product of two fields in order to avoid losing information about real correlations between the galaxy overdensity and the contaminants (see Eq.\ \ref{eq:a_coeff_gc}). 
    The choice of $\ell_{\rm max}$ depends on the contaminants in question. To determine a suitable value for the templates used here, we rerun the suite of simulations using several different values of $\ell_{\rm max}$ ranging from $\ell_{\rm max} = 10$ to $\ell_{\rm max} = 3N_{\rm side} - 1 = 767$\footnote{{\tt HEALPix} maps generated with a resolution parameter of $N_{\rm side}$ are approximately band-limited to multipoles with $\ell\lesssim 3N_{\rm side}-1$.}.

    \begin{figure*}
      \includegraphics[width=0.95\textwidth]{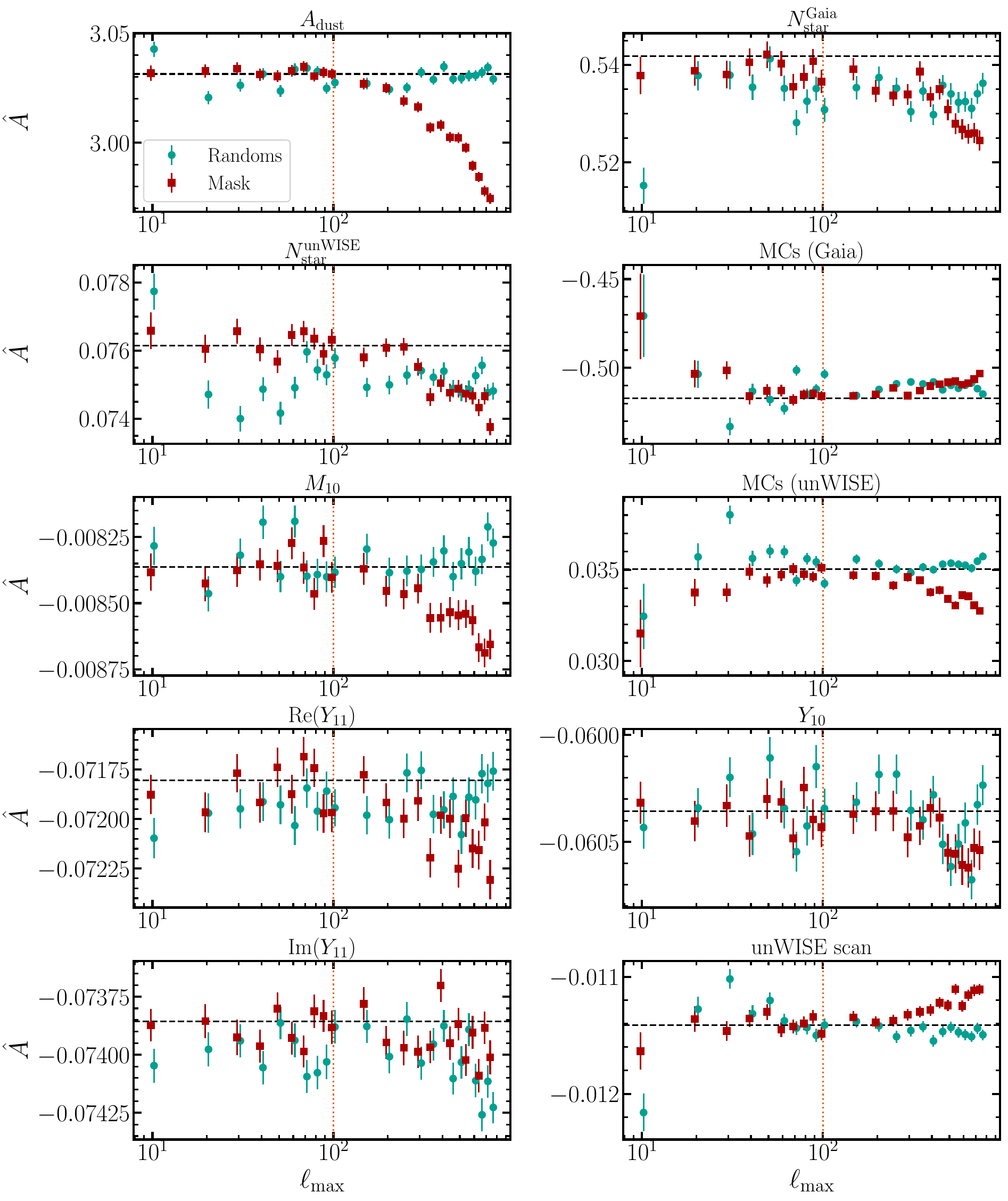}
      \caption{Linear deprojection coefficients measured as a function of $\ell_{\rm max}$ from simulations using the randoms-based (teal circles) and mask-based (red squares) approaches. The procedure is outlined in Section \ref{ssec:res.gcval}, and details of the templates (indicated at the top of each panel) can be found in Appendix \ref{sec:appx.templates}. The dashed black line in each panel shows the input deprojection coefficient used to contaminate the simulated data. Small horizontal offsets have been applied to the data for clarity. Both methods show significant variation in the deprojection coefficients for several contaminants, particularly as one moves to the lowest and highest multipoles. The dotted orange line indicates $\ell_{\rm max} = 100$, which we deem a suitable value for minimising the bias in our measured deprojection coefficients for the majority of templates.}
      \label{fig:alphas_vs_lmax}
    \end{figure*}

    The results of this test are shown in Fig.\ \ref{fig:alphas_vs_lmax}, wherein each panel shows the means and standard deviations in the deprojection coefficient $\hat{A}_p$ for a given contaminant as measured using the randoms- and mask-based approaches (teal circles and red squares, respectively), along with the input value used when generating the simulated data. Both the randoms- and mask-based deprojection methods produce biased estimates of the deprojection coefficients for several of the contaminants evaluated here if one uses an $\ell_{\rm max}$ that is either too low or too high. 
    We find that the most stable and least biased estimates tend to occur at $\ell_{\rm max} \sim 100$, and as such adopt this value going forward with our simulations.

    The reason for this bias in the estimated $\hat{A}_p$ is related to the pixel window function, and is a by-product of the fact that our contaminant templates take the form of maps. While in the map-based formalism it is straightforward to correct for the pixel window function, we can see from Eq.\ \ref{eq:a_coeff_gc} that such a correction is not applicable here, owing to calculation of ${\rm PCL}_\ell(a^v,\tilde{f}^q)$ required to obtain $\hat{A}_p$. We reiterate that $a^v$ is the (masked) catalogue-based field being observed, and the $\tilde{f}^q$ is a given template (also masked), the nature of which depends on which approach is being used. In the mask-based approach, the template maps are used directly, resulting in a PCL cross-correlating a catalogue-based field with a map-based field, for which a correction for the pixel window function is not viable. One might think this is mitigated by using the randoms-based approach, in which $\tilde{f}^q$ is computed by sampling the template at the positions of the randoms, but since these values are still being sampled from a map of finite resolution, the pixel window function will still have an effect that is not easy to remove. 
    While this effect is inherent to our simulations, it is also likely in a realistic scenario that one will be working with pixelised contaminant templates, in which case this will still be an issue. Since systematics tend to affect clustering measurements at larger scales, a relatively conservative $\ell_{\rm max}$ of ${\sim}100$ should be sufficient to capture the contribution from these sytematics in most use cases, while significantly mitigating the effects of the pixel window function.

    \begin{figure}
      \includegraphics[width=0.98\columnwidth]{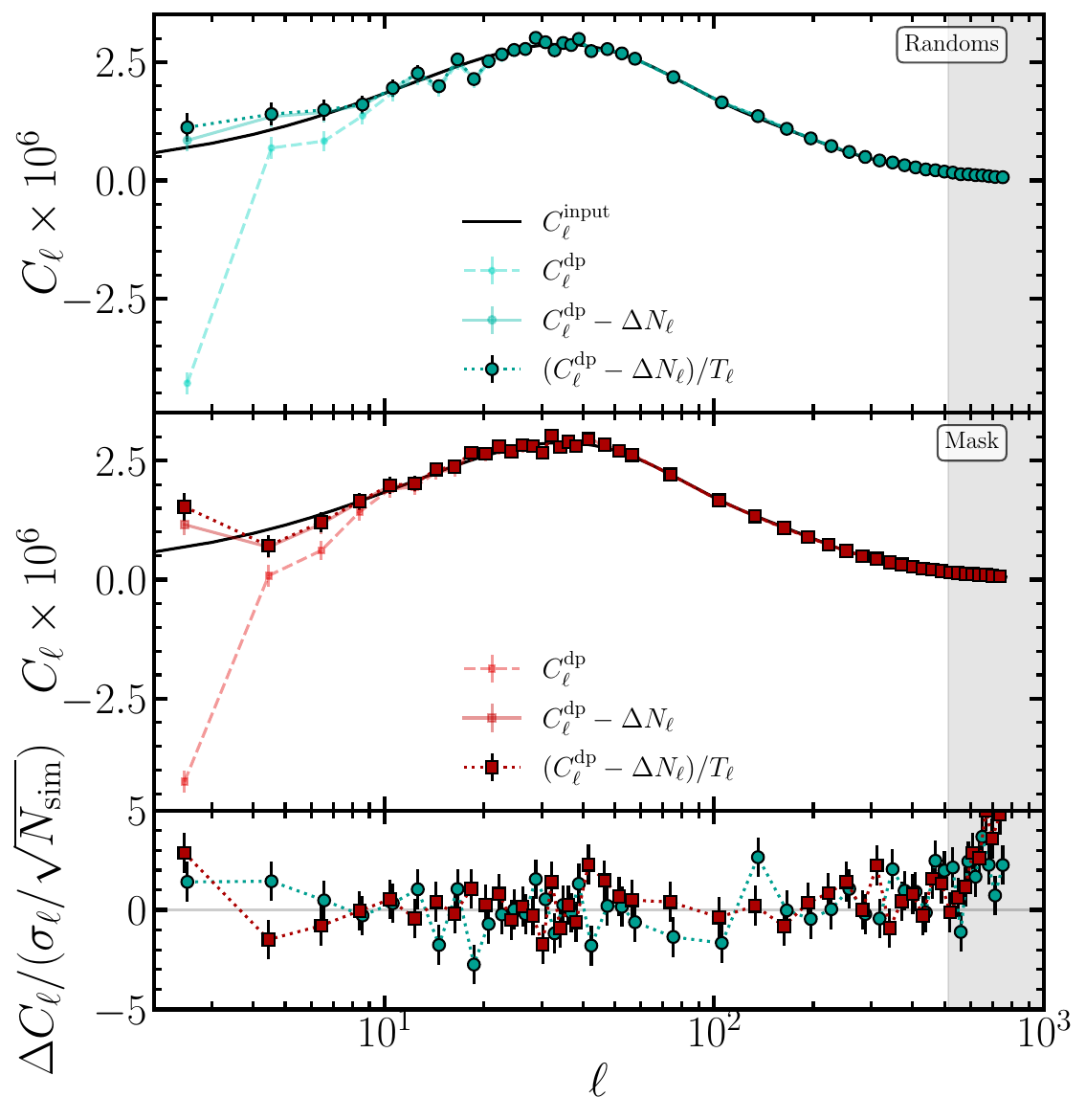}
      \caption{
        Means and standard deviations in angular power spectra from contaminated simulations measured at various stages of correction using the randoms-based (top panel) and mask-based (middle panel) approaches. Different marker sizes and linestyles represent the different stages: after contaminant deprojection (small markers, dashed lines); after removing the noise deprojection bias (medium-sized markers, solid lines); after also correcting for deprojection bias in the signal component using the transfer function approach (large markers, dotted lines). The black solid line shows the angular power spectrum used to generate the (non-contaminated) data, and as such is indicative of the `true' power spectrum. The shaded region begins at $\ell = 2N_{\rm side}$, beyond which we expect aliasing effects to arise from the fact that we have sampled the catalogue from a map with finite resolution. The bottom panel shows the residuals of the fully corrected power spectra with respect to the input power spectrum, normalised by the error in the mean across the simulations. 
        Both the randoms-based (teal circles) and mask-based (red squares) approaches are able to accurately reproduce the input power spectrum, with the randoms-based approach maintaining better accuracy than the mask-based at the higher multipoles. 
        Exacerbated deviations from the input power spectrum after applying the transfer function ($T_\ell$) at the largest scales are due to the data being noise-dominated in this regime (see text).
        }
      \label{fig:cl_rand_vs_mask}
    \end{figure}

    Figure \ref{fig:cl_rand_vs_mask} shows the result of applying the template deprojection procedure outlined in Section \ref{sssec:meth.deproj.gc} to the set of contaminated catalogues from our simulations, using both the randoms-based (teal circles; top panel) and mask-based (red squares; middle panel) approaches. 
    As was seen for sampled fields in Section \ref{ssec:res.sampval}, there is a significant loss of power at large scales, which is almost entirely mitigated by accounting for the noise deprojection bias via Eq.\ \ref{eq:deltaN_gc}, since these data are noise-dominated in this regime. 
    Consequently, attempts to apply the transfer function approach outlined in Section \ref{ssec:meth.dpbias} appear to exacerbate the deviations from the true power spectrum, although these shifts are marginal and the resulting power spectra are still consistent within the errorbars. 
    Overall, the final $C_\ell$ estimates are consistent with the input power spectrum to within $\sim3\sigma$ at all scales. Since our simulated catalogues have been sampled from maps of finite resolution, aliasing effects are expected to arise beyond $\ell \approx 2N_{\rm side}$; this can be seen for the mask-based case in the shaded grey region of the bottom panel in Fig. \ref{fig:cl_rand_vs_mask},while the randoms-based approach appears relatively unaffected. We therefore caution users against using the mask-based approach to probe scales of $\ell > 2 N_{\rm side}$.
    Importantly, we also note that the uncertainties in Fig. \ref{fig:cl_rand_vs_mask} correspond to the mean over 1000 simulated datasets, and are therefore $\sqrt{1000}\sim30$ times smaller than the actual errors for a real-world dataset. The small deviations shown in the figure should therefore be completely negligible in a realistic scenario.

  \subsection{Comparison between debiasing methods}\label{ssec:res.db_vs_tf}
    We now use simulations to validate the transfer function approach described in Section \ref{ssec:meth.dpbias} (see Eq. \ref{eq:transfer_func}) as a means of correcting for mode loss when deprojecting templates, and compare with the deprojection bias approach normally employed for this purpose (see Eq. \ref{eq:dpbias}; \citealt{1992ApJ...398..169R, 1609.03577}). In particular, we wish to test how dependent the transfer function approach is on assumptions about the true power spectrum.

    As in Section \ref{ssec:res.gcval}, we begin with 1000 Gaussian realisations of the angular power spectrum obtained from the {\sl Quaia} survey, where we have contaminated each realisation with those same templates via Eq.\ \ref{eq:dg_cont} and applied the {\sl Quaia} mask $w$. We will refer to this as the `main' suite of simulations henceforth. Using the standard map-based PCL approach (Sections \ref{sssec:meth.pcl.pcl} and \ref{sssec:meth.pcl.dpj}), we then deproject the templates to obtain biased estimates of the input power spectrum, $\hat{C}_\ell$. The transfer function or the deprojection bias estimates are  then used to debias these angular power spectra. We use the debiased power spectra as a means of assessing the efficacy of the transfer function approach.
    
    To calculate the various transfer functions we wish to test, we generate several more suites of simulated maps, each producing 1000 Gaussian realisations of a different input $C_\ell$. For each realisation, we apply the same mask as for the main suite and use the standard PCL approach to measure the angular power spectrum of the masked field, $C_\ell^{\rm before}$. We then use the map-based approach to deproject the same templates used to contaminate the main suite and measure the angular power spectrum post-deprojection, $C_\ell^{\rm after}$. The transfer function is then computed using Eq.\ \ref{eq:transfer_func}, averaging the relevant quantities across all simulations in the suite. 

    We use the following angular power spectra as inputs for each simulation suite:
    \begin{itemize}
      \item The {\sl Quaia} $C_\ell$ used to generate the the main suite, whose measured angular power spectra we aim to debias. This is the true power spectrum and therefore should, by definition, produce unbiased results after applying the transfer function.
      \item A guess at the true power spectrum, derived by measuring the PCL of one simulated {\sl Quaia}-like dataset and dividing by the covered sky fraction, estimated as $f_{\rm sky}=\langle w^2\rangle$.
      \item A power spectrum with an arbitrarily chosen functional form, $C_\ell = 1 / (\ell + 10)$.
      \item A flat power spectrum with arbitrary amplitude, $C_\ell = K$.
    \end{itemize}
    In addition, we use the standard deprojection bias approach to debias the measured power spectra, using one of the first two $C_\ell$s described above as the estimate of the true power spectrum ($C_\ell^{\rm true}$ in Eq. \ref{eq:dp.f.aux}).

    \begin{figure}
      \includegraphics[width=0.95\columnwidth]{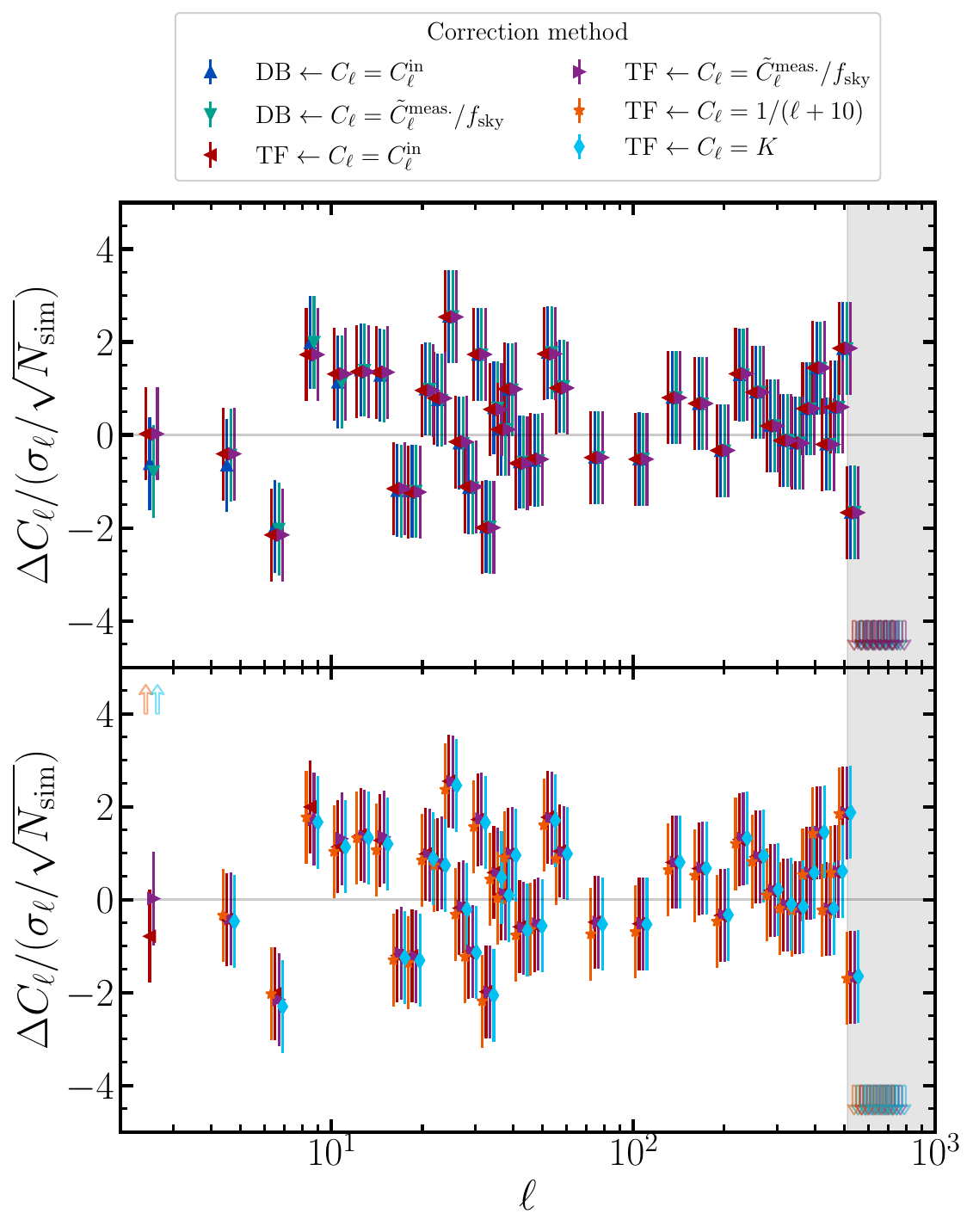}
      \caption{
        Residuals of the angular power spectra measured from simulations with respect to the power spectrum used to generate the simulated data, having mitigated the mode loss due to deprojection using different approaches. Namely, either the standard deprojection bias (DB) approach or the transfer function (TF) approach is used, with the chosen guess at the input power spectrum indicated in the legend. Small horizontal shifts have been applied to the data to aid visualisation, and the results have been partially split between two panels for ease of comparison between the most relevant methods. Arrows of a given colour denote scales at which the angular power spectrum deviates from the truth by more than $5\sigma$ after debiasing with the corresponding method. 
        {\it Top panel}: The DB approach using the true angular power spectrum (blue upward-pointing triangles) and a data-driven guess at its form (teal downward-pointing triangles), along with the TF approach using the same angular power spectra (red left-pointing and purple right-pointing triangles, respectively). The DB and TF approaches produce similarly unbiased $C_\ell$ estimates in both cases.  
        {\it Bottom panel}: The TF approach using the four different angular power spectra described in the text and indicated in the legend. 
        All four perform similarly well overall, although the choice of guess $C_\ell$ matters more at low $\ell$.
        }
      \label{fig:db_vs_tf}
    \end{figure}

    Figure \ref{fig:db_vs_tf} compares the deviation (normalised by the standard error of the mean) of the measured angular power spectra from the input power spectrum, after correcting for mode loss due to deprojection using one of the methods outlined above. We find that all methods perform similarly well at most scales, being consistent with each other within the errorbars. 
    Ignoring scales at which $\ell \gtrsim 2N_{\rm side}$ (demarcated by the grey shaded region in the figure) where aliasing effects are expected to arise from the pixelisation of our maps, it is only at the largest scales ($\ell \lesssim 3$) that we see any difference between the various methods. Namely, the transfer function approach fails to produced unbiased estimates at these scales if an unrealistic guess at the true angular power spectrum is used (see orange stars and cyan diamonds in Fig. \ref{fig:db_vs_tf}). 
    We thus conclude that the transfer function approach is a viable alternative to calculating the deprojection bias, so long as one uses a reasonable guess at the true power spectrum when generating the required simulations. For a realistic scenario in which the truth is not known, one can derive a sufficient guess by dividing the measured PCL by the observed sky fraction $f_{\rm sky}$, as has been done here (teal downward-pointing and purple right-pointing triangles in Fig. \ref{fig:db_vs_tf}, for the deprojection bias and transfer function approaches, respectively).

    \subsection{Catalogue-based transfer functions}\label{ssec:res.cat_tfs}
    While we find in Section \ref{ssec:res.db_vs_tf} that the approach of using map-based simulations to debias catalogue-based data appears to work sufficiently well, one question that remains is whether one can instead use catalogue-based simulations such as the those used in Section \ref{ssec:res.gcval} to estimate a transfer function that more fairly represents the data in question. 

    To test this, we again begin by generating 1000 Gaussian realisations of an input $C_\ell$. Since we find that the transfer function derived using a data-driven guess for the true $C_\ell$ performs better than if some arbitrary $C_\ell$ is used (see Figure \ref{fig:db_vs_tf}), we opt for this as our input $C_\ell$ here. This also represents a more realistic scenario one would encounter when working with real data, for which the true power spectrum is unknown. From each of these Gaussian realisations we then draw Poisson realisations of a catalogue with a mean of 644786 sources per catalogue (i.e. the same number of sources as in the {\sl Quaia} `low-$z$' sample), following the procedure outlined in Section \ref{ssec:res.gcval}. From these we compute the angular power spectra using the catalogue-based approach, both with and without deprojecting the contaminant templates. We do this for both the case where the mask is provided directly as a continuous map, and the case where randoms are used to define the mask, and compute the transfer function for each using Eq. \ref{eq:transfer_func}. 

    In both cases, we find that the catalogue-based simulations produce significantly noisier transfer functions than map-based simulations using the same input $C_\ell$, with the scatter across simulations being up to $\sim$ 100 (1000) times greater for the randoms-based (mask-based) approach. This is due to the impact of shot noise on the measured spectra before and afeter deprojection. We therefore recommend that the map-based approach of Section \ref{ssec:res.db_vs_tf} be used to derive transfer functions, particularly when working with shot noise-dominated datasets. Note that this also has the advantage of being less expensive computationally in most real-world applications.

  \subsection{Application to Quaia}\label{ssec:res.data}
    Having validated our method using simulations, we now test its application to real data. We again turn to the {\sl Quaia} catalogue \citep{2306.17749}, focusing in particular on the `low-$z$' sample of 644768 quasars with spectro-photometric redshifts in the range $z_p < 1.47$. 
    After removing any quasars with redshift estimates outside of this range, we deproject the same templates used by \cite{2410.24134,2504.20992} (see Appendix \ref{sec:appx.templates} for details) to mitigate the contribution of systematics. We do this using three approaches for the sake of comparison: the standard map-based approach, the catalogue-based approach in the case of a continuous mask, and the catalogue-based approach using randoms to define the mask. 
    In the latter two approaches, the shot noise is calculated with Eq. \ref{eq:cat_noise} (ignoring the contribution from randoms in the mask-based case) and subtracted from the measured power spectra. 
    For the map-based approach, we assume that the shot noise is entirely Poissonian, in which case it can also be calculated analytically \citep{1809.09603, 2020JCAP...03..044N}: the PCL of the noise is given by 
    \begin{equation}\label{eq:pnl}
      \tilde{N}_\ell = \Omega_{\rm pix} \frac{\langle w \rangle}{\langle n \rangle}\,,
    \end{equation}
    where $\Omega_{\rm pix}$ is the pixel area in steradians ($\approx 1.6 \times 10^{-5}$~sr for $N_{\rm side} = 256$), $\langle w \rangle$ is the mean value of the {\sl Quaia} mask across the map, and $\langle n \rangle$ is the mean effective number of sources per pixel. 
    The power spectrum of the shot noise is then obtained by multiplying it by the inverse mode-coupling matrix calculated from the {\sl Quaia} mask. 

     \begin{figure}
      \includegraphics[width=0.95\columnwidth]{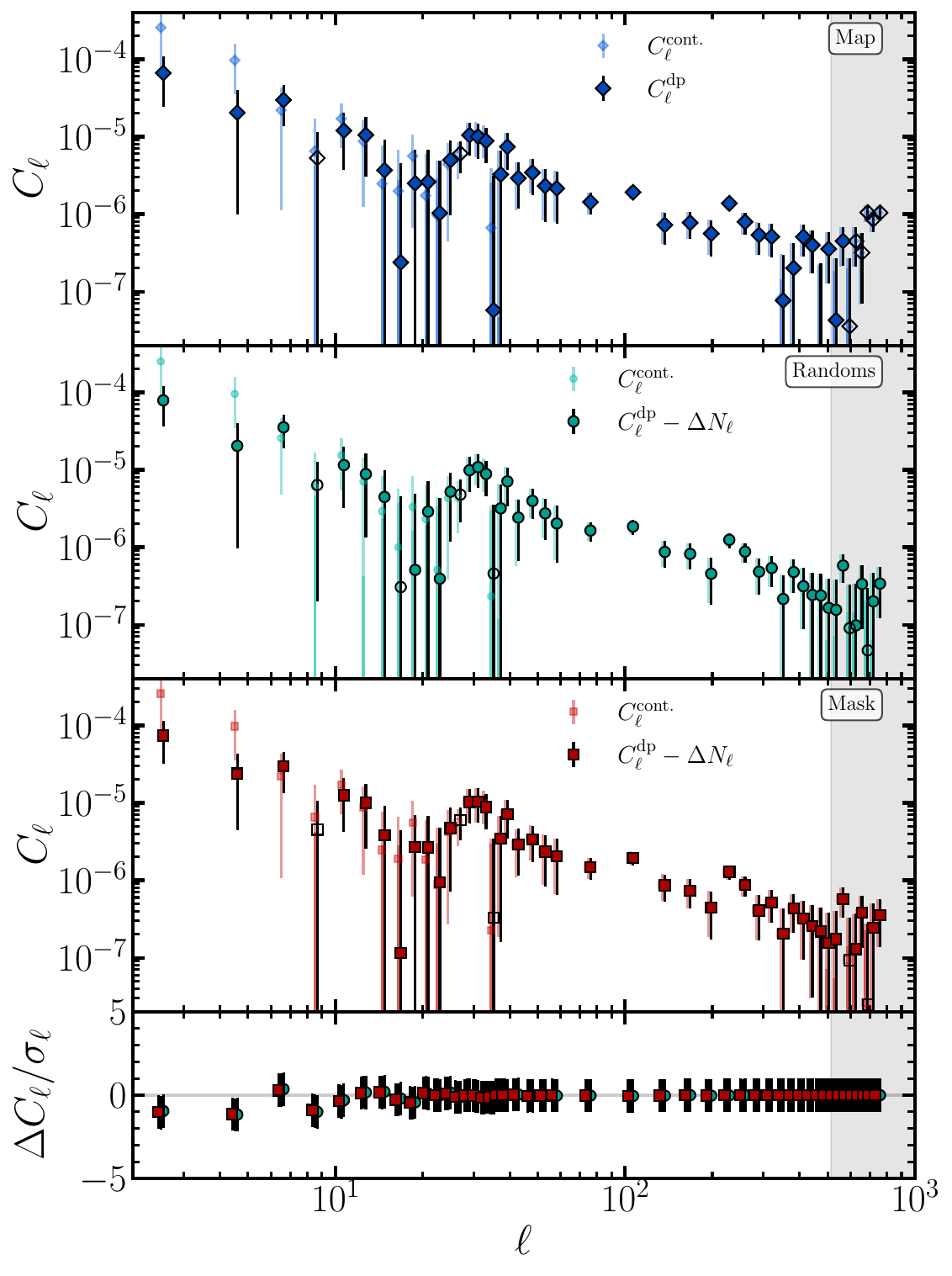}
      \caption{
        Application of three methods of deprojection to the low-$z$ sample of quasars from {\sl Quaia}. The first three panels show (from top to bottom) the results of applying the traditional map-based approach to template deprojection, the catalogue-based approach in the case where randoms have been used to define the mask, and the catalogue-based approach where the mask has been provided as a continuous map. In each case, we show the angular power spectra pre- and post-deprojection (correcting for the noise deprojection bias in the catalogue-based approaches). The shot noise has already been subtracted in all cases. Open symbols denote negative power, for which we plot the absolute values. The bottom panel shows the difference between the pre- and post-deprojection angular power spectra, normalised by the corresponding uncertainties. All three methods are found to result in similar power deviations across all scales.
        }
      \label{fig:quaia_deproj}
    \end{figure}

  Figure \ref{fig:quaia_deproj} shows the noise-subtracted angular power spectra pre- and post-deprojection as measured with the three different approaches. Uncertainties are estimated via the computation of Gaussian (or `disconnected') power spectrum covariance matrices. 
  Once the noise deprojection bias has been accounted for in the catalogue-based approaches, all three methods produce very similar measurements with respect to each other, both before and after template deprojection has been applied. 
  The consistency of the deprojection methods themselves is exemplified in the bottom panel, in which we show the deviation in power induced by this step, normalised by the uncertainties: the deviations agree within errorbars for all three methods at all angular scales. 
  This test thus validates our implementation of template deprojection in the catalogue-based PCL framework when applied to real data.

\section{Conclusions}\label{sec:conc}
  In this work, we present an extension to the catalogue-based PCL formalism of \cite{2407.21013} that allows one to apply template deprojection in order to mitigate the contribution from systematic effects. In doing so, we combine the advantages of the existing catalogue-based formalism -- i.e. avoiding biases stemming from the pixelisation of the observed fields -- with systematics mitigation capabilities previously only available in the map-based formalism. We describe the application of this formalism to fields sampled at the positions of discrete sources, as well as to the particular case of galaxy clustering, in which the density of sources is the field of interest. Note that while the fields considered here are all scalar, our implementation is applicable to fields of arbitrary spin. A key part of this formalism lies in being able to account for mode loss in the noise component, induced by the deprojection procedure, which we demonstrate can be accurately computed analytically. We also introduce an alternative method to account for deprojection bias in the signal component, via the computation of a transfer function, and demonstrate its efficacy using simulations.

  For the case of discretely-sampled fields, we validate our implementation by simulating 1000 Gaussian realisations of an arbitrary angular power spectrum, which are then contaminated with 100 synthetic templates and subsequently Poisson-sampled at the positions of galaxies from a real dataset ({\sl Quaia}) to create 1000 realisations of a catalogue with a realistic source distribution. We show that our implementation can be used to obtain unbiased estimates of the true power spectrum, after the mode loss induced by deprojection has been accounted for in both the noise and signal components.

  We then simulate 1000 {\sl Quaia}-like catalogues and test the clustering-specific part of our formalism. We first demonstrate that care must be taken when choosing a maximum multipole, $\ell_{\rm max}$, for the deprojection procedure, as choosing a scale that is too small or too large can result in biased estimates of the deprojection parameters (which in turn will bias the final angular power spectrum). Having selected a suitable multipole ($\ell_{\rm max} = 100)$, we show that unbiased estimates of the true power spectrum can again be obtained after correcting for mode loss in the noise component. Since these simulated catalogues are shot noise-dominated, the effect of the transfer function is minimal in comparison, but we emphasise that for denser datasets this step will likely be necessary to achieve unbiased results. We find that all of this is true regardless of whether the mask is defined via a catalogue of randoms, or as a continuous map.

  To assess the efficacy of the transfer function approach to mitigating mode loss caused by deprojection, we derive four transfer functions using different $C_\ell$s as input for the simulations, and compare their performance with the standard deprojection bias approach by applying them each to a set of 1000 simulated {\sl Quaia}-like catalogues. All methods are found to produce similarly unbiased results, except at scales $\ell \lesssim 3$, where at least a reasonable approximation of the true angular power spectrum is required.

  Finally, we apply our formalism to real data from {\sl Quaia} and compare it to the standard map-based approach. After correcting for the noise deprojection bias, both the randoms-based and mask-based approaches produce measurements consistent with the map-based approach, pre- and post-deprojection.

  We have thus demonstrated that our formalism is a viable alternative to the traditional map-based approach when working with fields sampled at discrete positions. We anticipate that the method presented here will be useful in the analysis of large datasets incoming from Stage IV surveys such as the Vera C. Rubin Observatory's Legacy Survey of Space and Time (LSST), for which systematics mitigation will be a crucial step towards obtaining unbiased cosmological measurements, and the wealth of information contained at small scales will be better exploited by avoiding pixelisation. Our implementation is made available with the latest public release of the public PCL package \nmt.

  \section*{Acknowledgements}
  This work has been supported by STFC funding for UK participation in LSST, through grant ST/X00127X/1. DA and TC acknowledge support from the Beecroft Trust. 
  BL is supported by the Royal Society through a University Research Fellowship. 
  KW acknowledges support from the Science and Technology Facilities Council (STFC) under grant ST/X006344/1. 
  Extensive use was made of computational resources at the University of Oxford Department of Physics, funded by the John Fell Oxford University Press Research Fund.

\section*{Data Availability}

The quasar catalogue used in this work is publicly available at \url{https://zenodo.org/records/12805938}.



\bibliographystyle{mnras}
\bibliography{main}



\appendix

\section{{\it Quaia} templates}\label{sec:appx.templates}

We include here brief details of the {\sl Quaia} systematics templates used for the analysis carried out in Sections \ref{ssec:res.gcval} and \ref{ssec:res.db_vs_tf}. 
These templates include: a Galactic dust extinction map ($A_{\rm dust}$) from \citet{2023ApJ...958..118C}; two maps of stellar contamination from {\sl Gaia} and {\sl unWISE} ($N_{\rm star}^{\rm Gaia}$ and $N_{\rm star}^{\rm unWISE}$, respectively); the scanning patterns of {\sl Gaia} and {\sl unWISE} ($M_{10}$ and {\it unWISE scan}); stellar density maps in the regions around the Large and Small Magellanic Clouds measured with {\sl Gaia} and {\sl unWISE} ({\it MCs (Gaia)} and {\it MCs  (unWISE)}); the three dipole moments associated with a multipole of $\ell = 1$ ($Y_{10}$, ${\rm Re}(Y_{11})$, ${\rm Im} (Y_{11}$)). 

\begin{figure*}
      \includegraphics[width=0.99\textwidth]{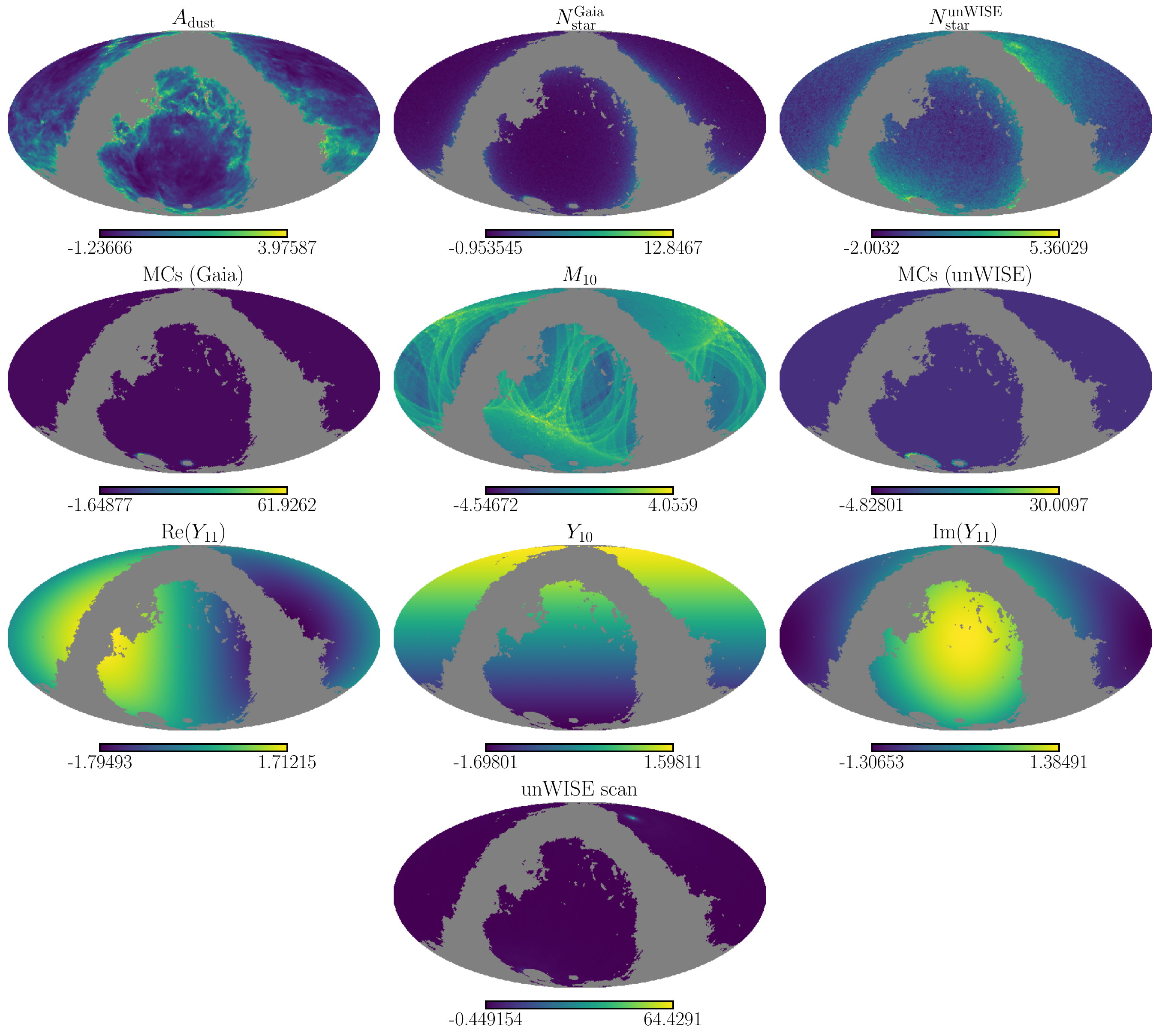}
      \caption{
        Systematic templates used when simulating clustering data in Sections \ref{ssec:res.gcval} and \ref{ssec:res.db_vs_tf}, and when applying our catalogue-based deprojection formalism to {\sl Quaia} data in Section \ref{ssec:res.data}. As in \citet{2504.20992}, we show for each systematic template $S(\nv)$ the quantity $(S(\nv) - \bar{S}) / \sigma_S$, where $\bar{S}$ and $\sigma_S$ are the mean and rms deviation within the footprint, respectively.
      }
      \label{fig:templates}
\end{figure*}

We show Mollweide-projected maps of these templates in Fig.\ \ref{fig:templates}. In each case, we show the residual of a given pixel with respect to the mean value of that systematic across the footprint ($S(\nv) - \bar{S}$), normalised by the rms deviation across the footprint ($\sigma_S$).

\bsp	
\label{lastpage}
\end{document}